\documentclass{sigchi-ext}
\usepackage[T1]{fontenc}
\usepackage{textcomp}
\usepackage[scaled=.92]{helvet} 
\usepackage{graphicx} 
\usepackage{balance}  
\usepackage{booktabs} 
\usepackage{ccicons}  
\usepackage{ragged2e} 

\def\plaintitle{SIGCHI Extended Abstracts Sample File: \underline{N}ote
  \underline{I}nitial \underline{C}aps} 
\def\emptyauthor{}
\def\plainkeywords{Virtual Reality; Navigation; Proxy Objects; High Fidelity Interaction}

\title{Virtual Smartphone: High Fidelity Interaction with Proxy Objects in Virtual Reality}

\numberofauthors{6}
\author{%
  \alignauthor{%
    \textbf{Gian-Luca Savino}\\
    \affaddr{University of Bremen} \\
    \affaddr{Bremen, Germany} \\
    \email{gsavino@uni-bremen.de} }\alignauthor{%
    }
    }

\definecolor{linkColor}{RGB}{6,125,233}
\hypersetup{%
  pdftitle={\plaintitle},
  pdfauthor={\emptyauthor},
  pdfkeywords={\plainkeywords},
  bookmarksnumbered,
  pdfstartview={FitH},
  colorlinks,
  citecolor=black,
  filecolor=black,
  linkcolor=black,
  urlcolor=linkColor,
  breaklinks=true,
}

\copyrightinfo{\scriptsize Permission to make digital or hard copies of part or all of this work for personal or classroom use is granted without fee provided that copies are not made or distributed for profit or commercial advantage and that copies bear this notice and the full citation on the first page. Copyrights for third-party components of this work must be honored. For all other uses, contact the owner/author(s). \\
{\emph{EPO4VR - 1st Workshop on Everyday Proxy Objects for Virtual Reality at CHI '20, April 26, 2020, Honolulu, HI, USA.}} \\
\copyright~2020 Copyright is held by the author/owner(s).}
\begin{document}

\CopyrightYear{2020}
\setcopyright{rightsretained}
\conferenceinfo{CHI'20,}{April  25--30, 2020, Honolulu, HI, USA}
\isbn{978-1-4503-6819-3/20/04}
\doi{https://doi.org/10.1145/3334480.XXXXXXX}

\maketitle

\RaggedRight{} 

\begin{abstract}
  This workshop paper presents two proxy objects for high fidelity interaction in virtual reality (VR): a paper map and a smartphone. We showcase how our virtual paper map can increase interactivity and orientation, while our virtual smartphone extends the use of a proxy object, as it allows for actual touch input on a real phone leading to an almost infinite set of possible (inter-)actions (e.g. snapping pictures in the virtual world). Observations showed that participants were very precise in holding and interacting with both the paper map and the smartphone even though they did not see their hands in VR. The interaction in general was very intuitive which was mostly attributed to the realistic size of the virtual objects.
  Using our findings we discuss the trade off between adaptivity and high fidelity of proxy objects in VR. 
\end{abstract}




\begin{figure*}[t]
  \centering
  \includegraphics[width=1\textwidth]{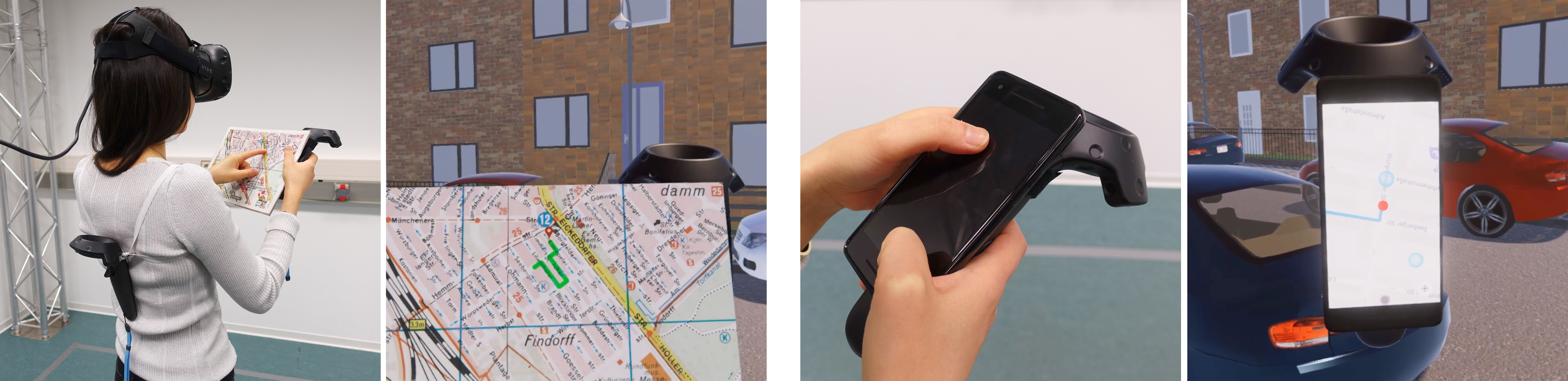}
  \caption{Left: Setup of the real and virtual paper map. Right: Setup of the real and virtual smartphone with visualised touch points.}~\label{fig:all_controllers}
\end{figure*}

\section{Introduction \& Motivation}
The lack of haptic feedback in virtual reality (VR) opened up a large area of research in HCI since the release of consumer HMDs like the HTC Vive and Occulus Rift. Many studies try to simulate haptic feedback by creating resistances~\cite{Zenner2019}, adding weight to the controllers~\cite{Zenner2017}, or providing surface textures~\cite{Whitmire2018}. This allows for multi-use and adaptive applications for a multitude of objects.

This flexibility also means that the haptic feedback can be limited and might not exactly represent the object of interest in its properties. To overcome these limitations, researchers have started to use proxy objects that represent their virtual counterpart in it's form, weight, texture, or all of them to provide the user with the most realistic and immersive feeling available~\cite{feick2020, Zhao2017, Azmandian2016}. This, of course, results in the loss of adaptivity and flexibility when it comes to the variety of objects each proxy can represent: A real apple only maps to a virtual apple and not much else. 

In this workshop paper we want to show that there are domain specific proxy objects that still retain some flexibility (within their domain application) while delivering a high fidelity interaction in VR. The following pages present two proxy objects: a paper map and a smartphone, which were both used as navigation aids in a VR study~\cite{Savino2019}. The aim of the study by Savino et al.~\cite{Savino2019} was to recreate an overall close to reality navigation experience in VR. To achieve this goal we developed high fidelity, interactive proxy objects that served as exact replicas of their virtual counterparts. 

As the above mentioned paper by Savino et al.~\cite{Savino2019} focuses mainly on the general comparison between a VR and real life experience, we want to take the opportunity in this workshop paper to highlight the functionality and setup of our high fidelity proxy objects. We discuss further use cases and showcase additional functionality that was not mentioned in the paper~\cite{Savino2019}.

\section{Technical Setup}
To properly interact with the virtual paper map and the virtual smartphone, their physical proxy objects need to be tracked alongside with the participants in VR. In our setup we used an additional HTC Vive controller that was attached to the paper map and the smartphone respectively using velcro (this was shortly before HTC released their separate trackers). The paper map supported attachment of a controller to either the left or the right side to account for handedness. The smartphone was placed within a rubber case, which was then attached to the controller (see figure \ref{fig:all_controllers}). As both hands were usually used to manipulate the proxy object, we used a walking-in-place locomotion method where participants wore a controller on their back (see figure \ref{fig:all_controllers}). The up-and-down movement was then translated into forward movement~\cite{Savino2019}.

We used a multiplayer architecture in Unity3D to send the touch input data from the Android phone to a server which hosted the virtual environment. This way the real smartphone just showed a black screen after starting the application but still sent all the touch input to the virtual smartphone. We then used it to classify different traditional touch gestures like tap, pinch and rotate. This allowed us, for example, to use well established gestures like pinch-to-zoom and two-finger-rotate in our mobile maps application as shown in figure \ref{fig:all_controllers}.

When creating the 3D versions of both the paper map and the smartphone, we made sure to use the exact dimensions of their real life counterparts. This proved to be very effective for the precision when interacting with the virtual devices.

\section{Observations}
While testing our setup as well as during the main study of~\cite{Savino2019} we made observations that we present and discuss in this section. From these we will extend further use cases and improvements. 

One interesting observation we made already during the testing phase was that the size of the virtual objects had a huge impact on the precision with which people were able to grasp the object. This supports the results by Kwon et al.~\cite{Kwon2009}, who found that an object was easier to grasp initially when the size of the proxy object and the virtual objects matched. It did not effect the manipulation after the initial grasp, though. In our case, once the paper map had the exact right dimensions, people were able to easily grasp it with two hands. In regards to the smartphone, this meant that once the size was right people were already very good at hitting the right buttons (e.g. the camera shutter) without any kind of additional feedback. To increase the accuracy for these kinds of tasks, we included markers on the screen that visualise the touch points (see figure \ref{fig:all_controllers}).

\begin{figure}[t]
  \centering
  \includegraphics[width=1\columnwidth]{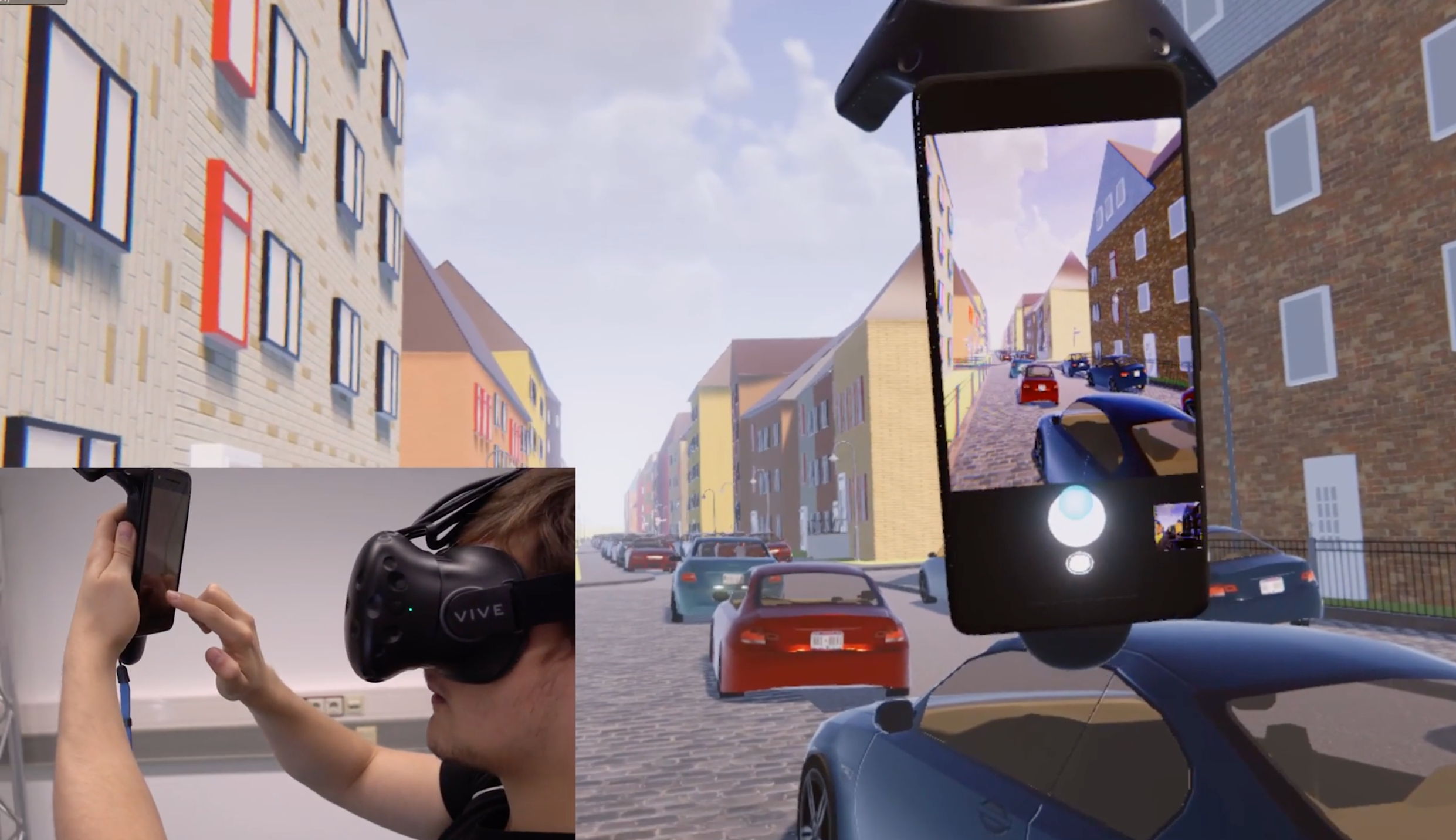}
  \caption{A participant using the camera application of our virtual smartphone.}~\label{fig:camera}
\end{figure}

\subsection{Using the Proxy Objects for Navigation}
In the case of a navigation task the interactive nature, good "graspability", and easy to hold controller of the paper map allowed for a very important action, namely rotating the paper map. As people have an easier time navigating when the map is in a track up alignment~\cite{Aretz1992} ("forward is up"), it is crucial that the map rotation is intuitive and comfortable. Using our proxy, participants were able to rotate the map successfully but were sometimes hindered by the one-sided weight distribution of the attached controller.

The smartphone, on the other side, did not suffer from unnatural weight distribution. Participants did however have problems manipulating the smartphone with a single hand, due to the added thickness from the controller. Since most gestures on mobile maps usually need two hands anyway, this wasn't much of a drawback. The two-finger-rotate gesture allowed for easy map manipulation and direction alignment. 

\section{Adaptability}
The navigation context really shows the adaptability of both proxy objects. The paper map, for example, was used for four different routes in the experiment. By just exchanging the visual representation of the virtual map we can use any kind of map section an reuse the proxy object in further experiments. It could even serve as any kind of text document in other contexts, for example as an information brochure about landmarks. 

The adaptability of the smartphone lies mainly it its feature to use any kind of application that is needed for any specific purpose. In the navigation context it allows for testing a multitude of established and novel navigation applications. But even outside of this scope, it allows for completely new interactions, e.g. taking smartphone pictures in VR like shown in figure \ref{fig:camera}.

\section{Conclusion}
In our study we experienced that participants were very precise at interacting with the virtual paper map and the virtual smartphone even though they were not able to see their own hands. Our observations show that, using proper tracking, as long as the virtual and the real objects share the same dimensions, it is rather easy to manipulate them in VR. Still, for future experiments, the tracking of participants' hands might increase their precision and will enable even more control when interacting with proxy objects.

We showcase how proxy objects that have a 1:1 mapping to a virtual object with all their properties can still be adaptively used in different contexts. Even more so, when there are domain specific needs that proxy objects need to fulfil, using exact replicas as proxy objects might be worth the trade-off between adaptability and high fidelity interaction.

\balance{} 

\bibliographystyle{SIGCHI-Reference-Format}
\bibliography{sample}

\end{document}